\documentclass[final,5p,times,twocolumn]{elsarticle}

\usepackage{graphicx}
\usepackage{epstopdf}
\usepackage{booktabs}

\usepackage{hyperref}

\usepackage{amsmath,amssymb}

\journal{Nuclear Instruments and Methods in Physics Research A }

\begin{document}

\begin{frontmatter}



\title{Beam Dynamics and Tolerance Studies of the THz-driven Electron Linac for the AXSIS Experiment}


\author[]{K. Galaydych\corref{m1}}
\cortext[m1]{Corresponding author}
\ead{kostyantyn.galaydych@desy.de}

\author{R. Assmann}
\author{U. Dorda}
\author{B. Marchetti}
\author{G. Vashchenko}
\author{I. Zagorodnov}

\address{DESY, Hamburg, Germany}

\begin{abstract}
A dielectric-loaded linac powered by THz-pulses is one of the key parts of the "Attosecond X-ray Science: Imaging and Spectroscopy" (AXSIS) project at DESY, Hamburg. As in conventional accelerators, the AXSIS linac is designed to have phase velocity equal to the speed of light which, in this case, is realized by tuning the thickness of the dielectric layer and the radius of the vacuum channel. Therefore, structure fabrication errors will lead to a change in the beam dynamics and beam quality. Additionally, errors in the bunch injection will also affect the acceleration process and can cause beam loss on the linac wall. This paper numerically investigates the process of electron beam acceleration in the AXSIS linac, taking into account the aforementioned errors. Particle tracking simulations were done using the code ECHO, which uses a low-dispersive algorithm for the field calculation and was specially adapted for the dielectric-loaded accelerating structures.
\end{abstract}


\end{frontmatter}


\vspace{50mm}

\section{Introduction}
The long term goal of the AXSIS project, which will be hosted at accelerator R\&D SINBAD facility (DESY, Hamburg), is to produce fully-coherent X-ray radiation in the attosecond range~\cite{Kartner_NIM_2016}. In order to generate this radiation, inverse Compton scattering of a laser pulse on an electron beam will be used. Beforehand, an electron beam will be accelerated up to about 20 MeV in the AXSIS linac in order to provide about 5 keV photons after inverse Compton scattering. To realize the acceleration, a cylindrical normal conducting waveguide loaded with dielectric material was chosen (see Fig.~\ref{Fig_1}). The first reason for this is that one of the eigenmodes of the accelerating structure is the fundamental azimuthal symmetric transverse magnetic ($\mathrm{TM_{01}}$) mode and an accelerating field in the form of this mode is desired. The AXSIS linac will be powered by a THz-pulse. Several pulse incoupling options are currently being investigated. One of these options is incoupling of a radially-polarized THz-pulse into the linac. Such a pulse will convert to the $\mathrm{TM_{01}}$-like mode field as was previously shown in~\cite{Nanni_Nature_2015}, and this is the second reason why we choose this type of the linac for the experiment. In this paper we assume this type of electromagnetic field excitation in the linac. Despite the simplicity of the accelerating structure and the working principle of the linac, there are some critical points, that dramatically affect the acceleration process and therefore are studied in detail. This include the accelerator fabrication errors and inaccuracy of the beam injection. This is the primary motivation of this work.
\section{Statement of the problem}
\label{sec:Introduction}
\begin{figure*}[!ht]
    \centering
    \includegraphics*[width=\textwidth]{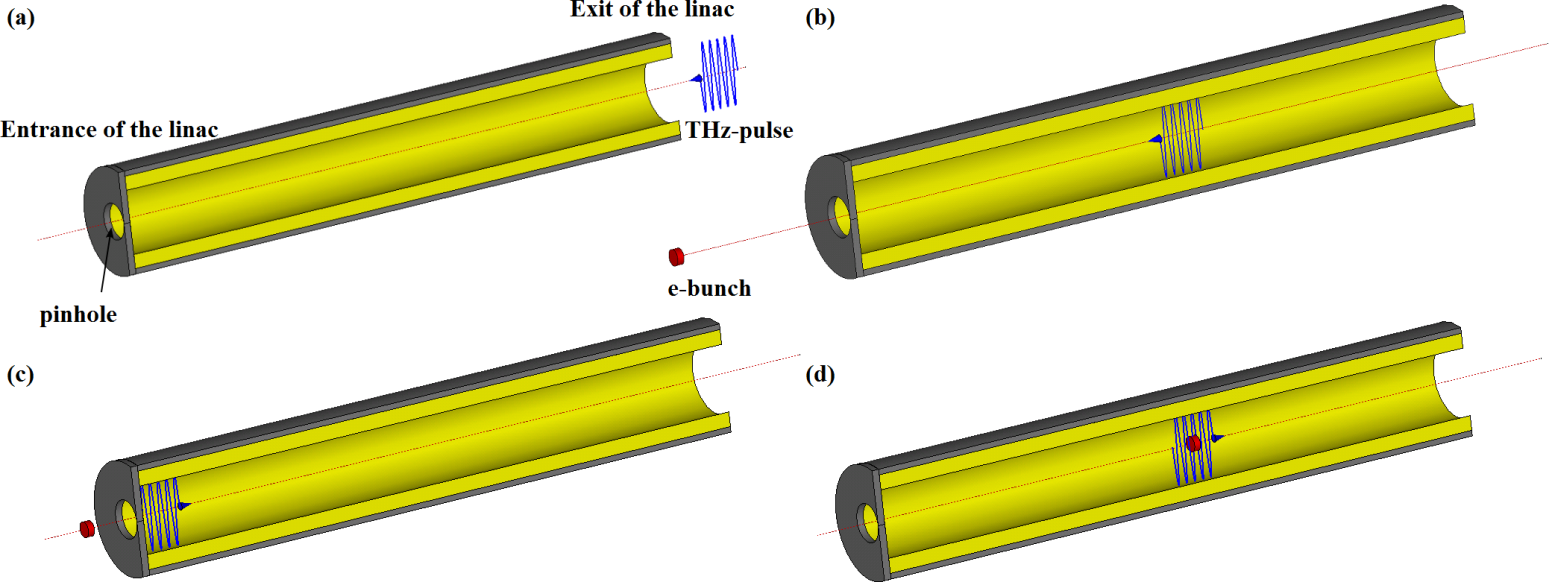}
    \caption{Overview of a dielectric loaded metal waveguide with an electron bunch and the metal plate with a pinhole aperture at the entrance of waveguide. The electron bunch moves from left to right. The working principle of the linac: a THz-pulse is focused onto the open end of the waveguide (a), propagates along the waveguide (b) and after reflection (c) co-propagates with the injected bunch (d).}
    \label{Fig_1}
\end{figure*}
Figure~\ref{Fig_1} demonstrates the working principle of the AXSIS linac, which is as follows: after focusing (omitted) and coupling (Fig.~\ref{Fig_1} (a)) the THz-pulse converts to the $\mathrm{TM_{01}}$-like mode field, propagates inside the linac (Fig.~\ref{Fig_1} (b)) and, after reflection (Fig.~\ref{Fig_1} (c)), propagates back with the injected electron beam downstream the linac (Fig.~\ref{Fig_1} (d)).

The main goal of this work is to determine the requirements for the production tolerances of the linac, based on simulation studies. We have investigated the influence of the linac manufacturing errors and the beam injection misalignment on the beam acceleration in terms of the maximum achievable output kinetic energy, the transverse and longitudinal phase space of the beam, the stability of the beam dynamics along the linac and the beam charge losses.

For the particle tracking simulations we used the ECHO code~\cite{Zagorodnov_Weiland_PRSTAB_2005,Zagorodnov_NAPAC_2016} which was specially adapted for dielectric-loaded accelerating structures of cylindrical and rectangular configurations. For the equations of particle motion, this code uses the standard leapfrog scheme.  For the calculation of the excited non-stationary electromagnetic field, a special low-dispersive algorithm was implemented in this code. This is done in order to avoid numerical dispersion on the spatial grid. In turn, it makes it possible to significantly improve the accuracy of calculation without increasing the grid density and avoiding nonphysical effects.

Our simulations were performed with parameters similar to the planned parameters of the linac and THz-pulse for the AXSIS project~\cite{Kartner_NIM_2016}. The parameters of the input bunch were taken from~\cite{Thomas_EAAC_2017} and are presented in Table~\ref{Tabl_1}. The parameters of the dielectric-loaded accelerator are chosen such that at the central frequency of the pulse, the phase velocity is exactly equal to the speed of light in vacuum.
\begin{table}[!hb]
  \centering
  \caption{Parameters of the linac, multi-cycle THz-pulse and electron bunch.}
  \label{tab:table1}
  \begin{tabular}{ccc}
    \toprule
    \textbf{Parameter} & \textbf{Value} & \textbf{Unit}\\
    \midrule
    Dielectric inner radius & 625 & $\mu m$\\
    Dielectric thickness & 77.0 & $\mu m$\\
    Rel. permittivity of the dielectric & 4.41 & \\
    Pinhole radius & 300 & $\mu m$\\
    Length of the linac & 100 & mm\\
    $\beta_{ph}$(300 GHz)/$\beta_{gr}$(300 GHz) & 1.0/0.62 & \\
    Beam charge & -1 & pC\\
    Average kinetic energy & 4.77 & MeV\\
    Energy spread & 10.21 & keV\\
    Horizontal (x) beam size, rms & 98.5 & $\mu m$\\
    Vertical (y) beam size, rms & 98.47 & $\mu m$\\
    Longitudinal (z) beam size, rms & 11.16 & $\mu m$\\
    Horizontal beam emittance & 0.38 & $\pi$ mm mrad\\
    Vertical beam emittance & 0.38 & $\pi$ mm mrad\\
    Longitudinal beam emittance & 5.36$\cdot10^{-2}$ & $\pi$ keV mrad\\
    Temporal shape of the pulse       & flat top & \\
    Central frequency of the pulse & 300 & GHz\\
    Duration of the pulse & 133.3 & ps\\
    Peak $E_z$ (on-axis) & 240 & MV/m\\
    \bottomrule
  \end{tabular}
  \label{Tabl_1}
\end{table}
\section{Simulation results}
\begin{figure}[!h]
  \centering
  \includegraphics[width=\linewidth]{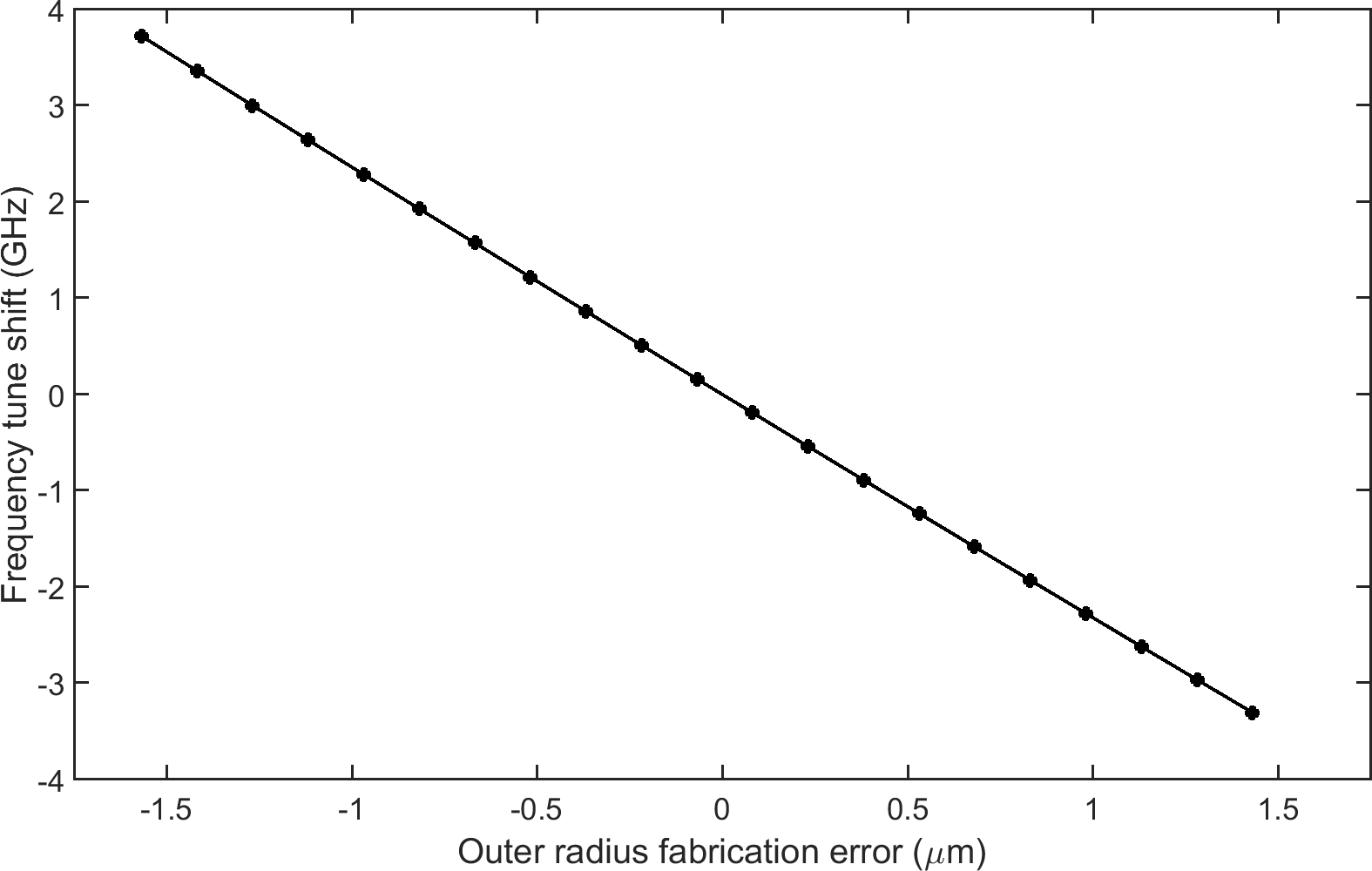}
  \caption{Dependence of the frequency correction on the outer dielectric radius fabrication error (designed dielectric thickness as in Table~\ref{Tabl_1}). The phase velocity equals the speed of light along the line.}
      \label{Fig_2}
\end{figure}

A phase velocity, especially for a THz-range, is extremely sensitive to the linac parameters and it could lead to undesired effects like phase slippage, and, as a result, to beam quality degradation. If the fabricated linac differs from designed one, then at the central frequency of the THz-pulse, the phase velocity does not equal to the speed of light due to frequency dispersion. We start with the case of the fabrication errors. We assume that we have fixed all the parameters except for the dielectric thickness. The natural way to fix this error is to tune the frequency to satisfy the desired phase velocity. The first question is how the frequency of the THz-pulse should be corrected in order to achieve (after changing the dielectric thickness) a phase velocity exactly equal to the speed of light at the new (corrected) frequency. Using a dispersion equation for TM modes of the dielectric loaded cylindrical waveguide~\cite{Bruck,Frankel} this problem was solved numerically. Figure~\ref{Fig_2} shows the result of this correction.

It can be seen that this dependence is linear and, for example, for the outer dielectric radius fabrication error of 1.5 $\mu m$ we should change the THz-pulse frequency from 300 GHz to 296.696 GHz. It means that the correction will need a frequency tuning precision of about 1 MHz.

\begin{figure}[!h]
  \centering
  \includegraphics[width=\linewidth]{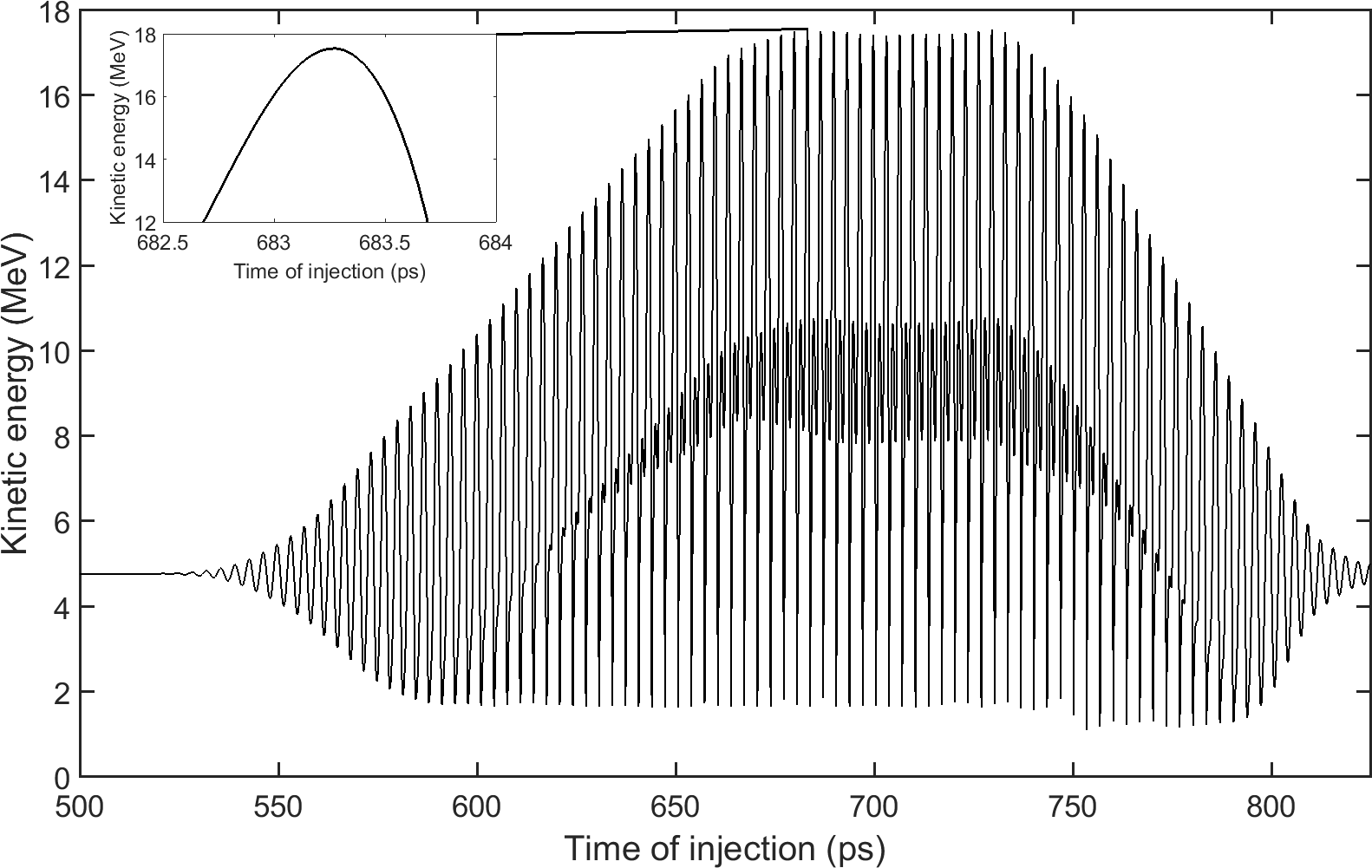}
  \\
  \includegraphics[width=\linewidth]{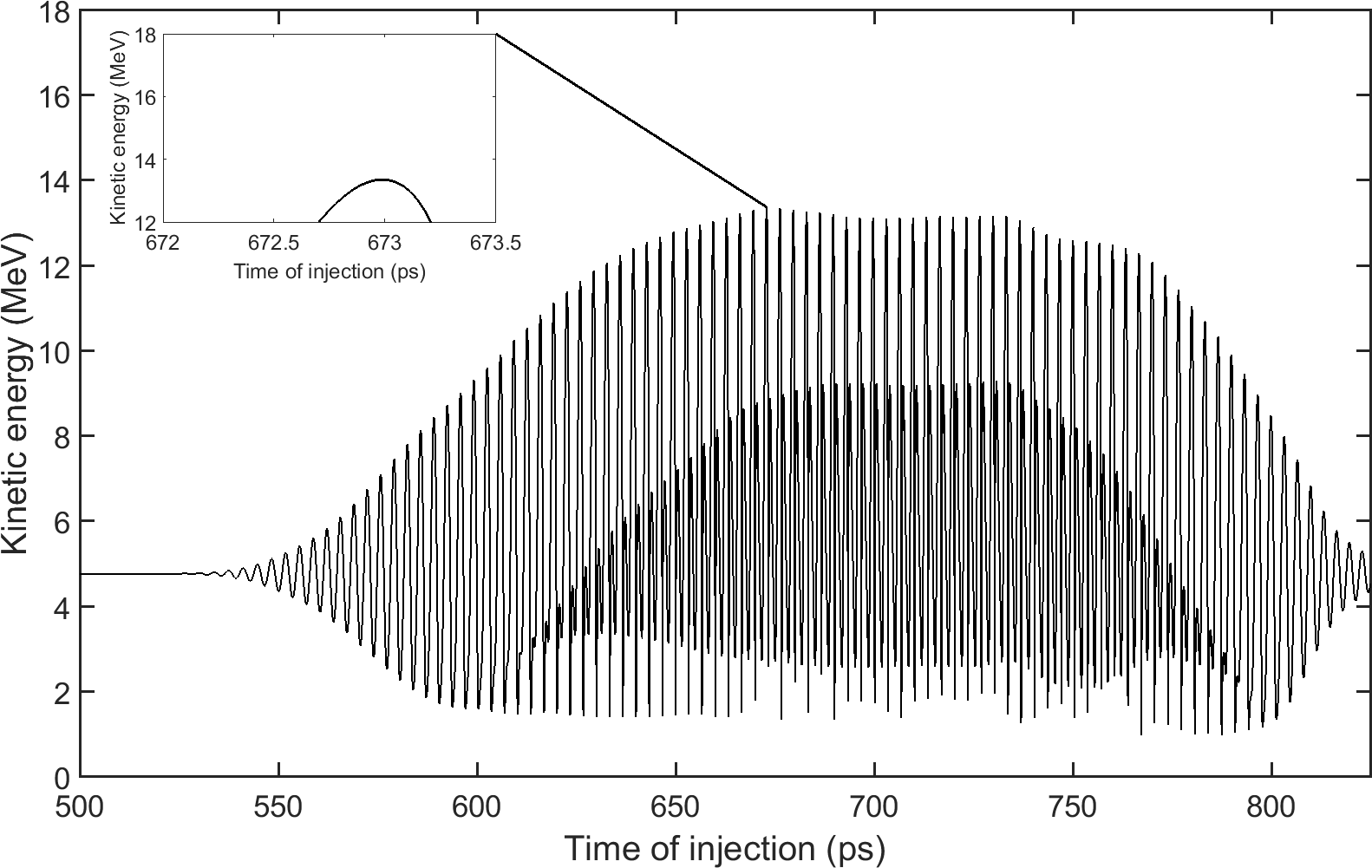}
  \\
  \includegraphics[width=\linewidth]{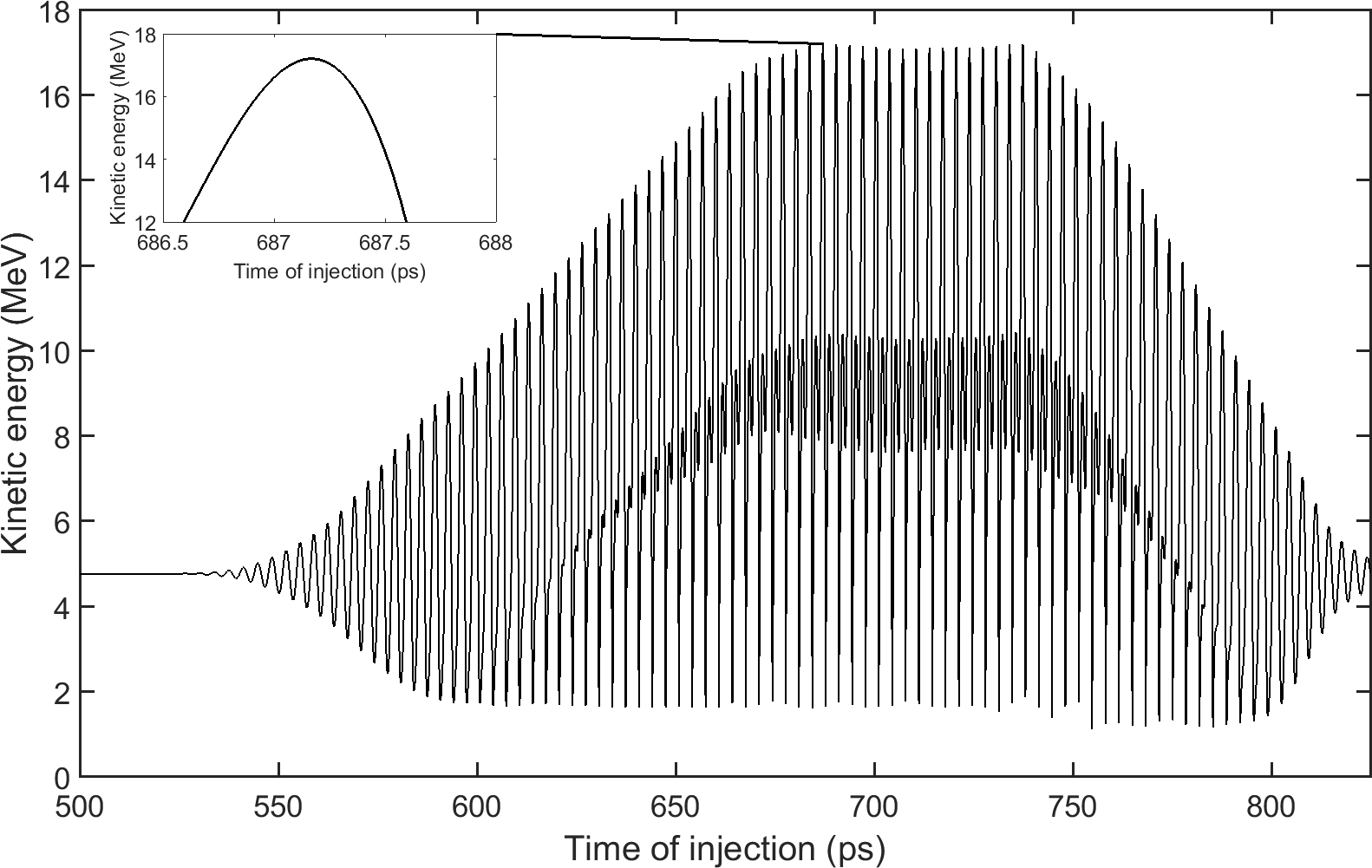}
  \caption{The output kinetic energy as a function of the injection time for the cases: (1, top) - optimized linac (with exact parameters); (2, middle) - detuned linac with 1.5 $\mu m$ outer dielectric radius fabrication error without the frequency correction; (3, bottom) - detuned linac with 1.5 $\mu m$ outer dielectric radius fabrication error with the frequency correction.}
      \label{Fig_3}
\end{figure}

\begin{figure}[!h]
  \centering
  \includegraphics[width=\linewidth]{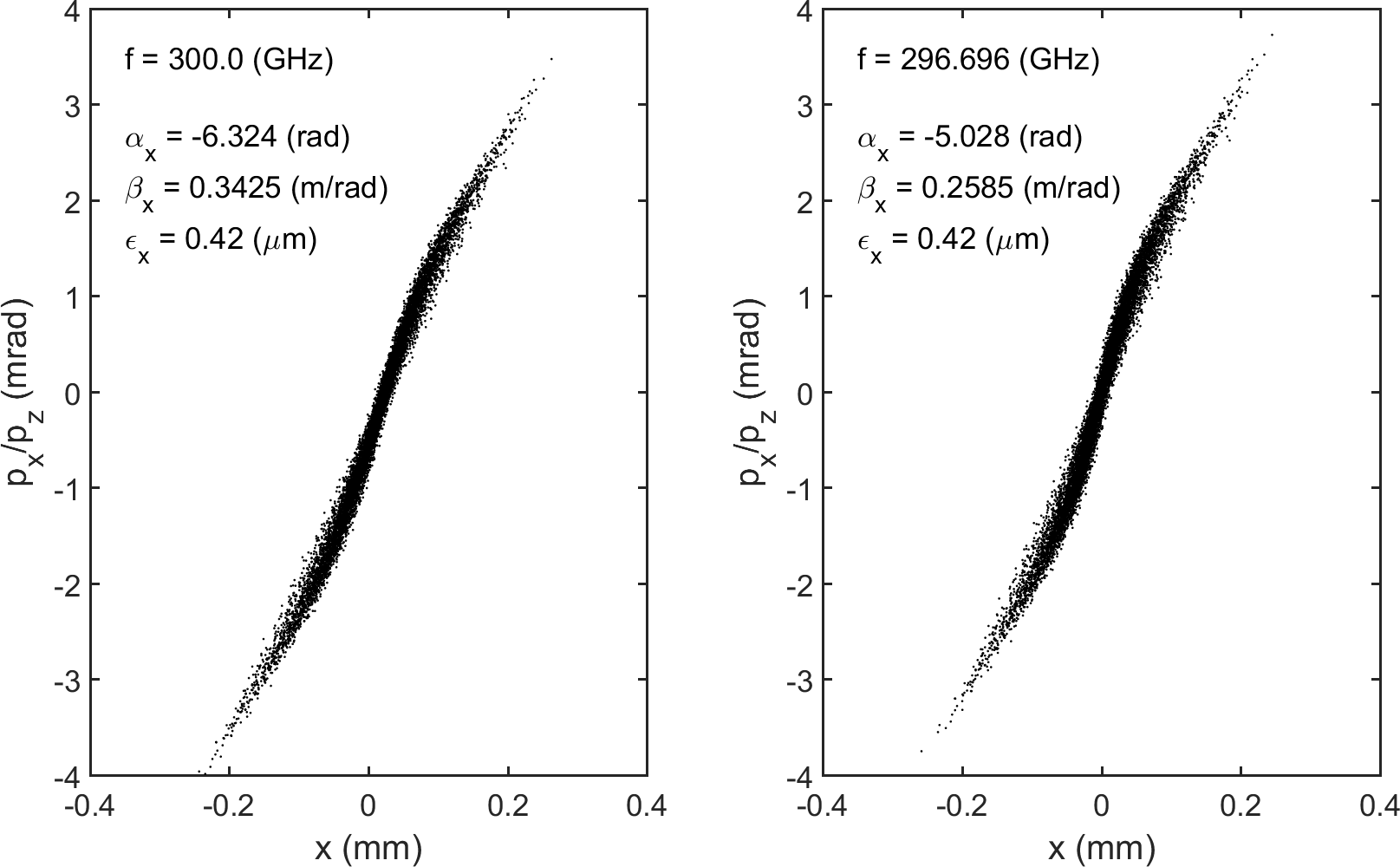}
  \\
  \includegraphics[width=\linewidth]{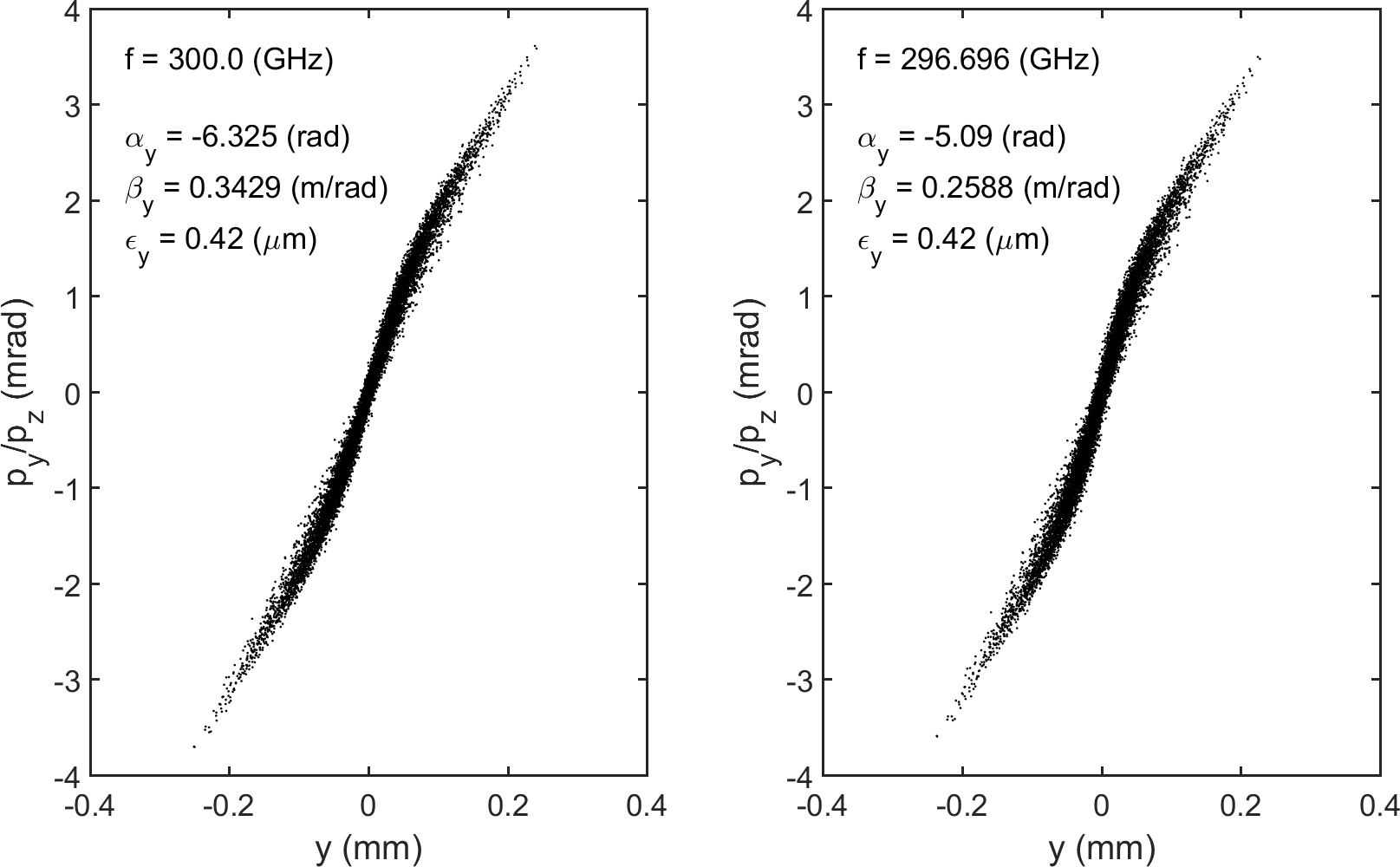}
  \caption{Transverse phase space of the bunch for the tuned linac (left column) and for the detuned linac with corrected frequency of the THz-pulse (right column). }
      \label{Fig_4}
\end{figure}

\begin{figure}[!h]
  \centering
  \includegraphics[width=\linewidth]{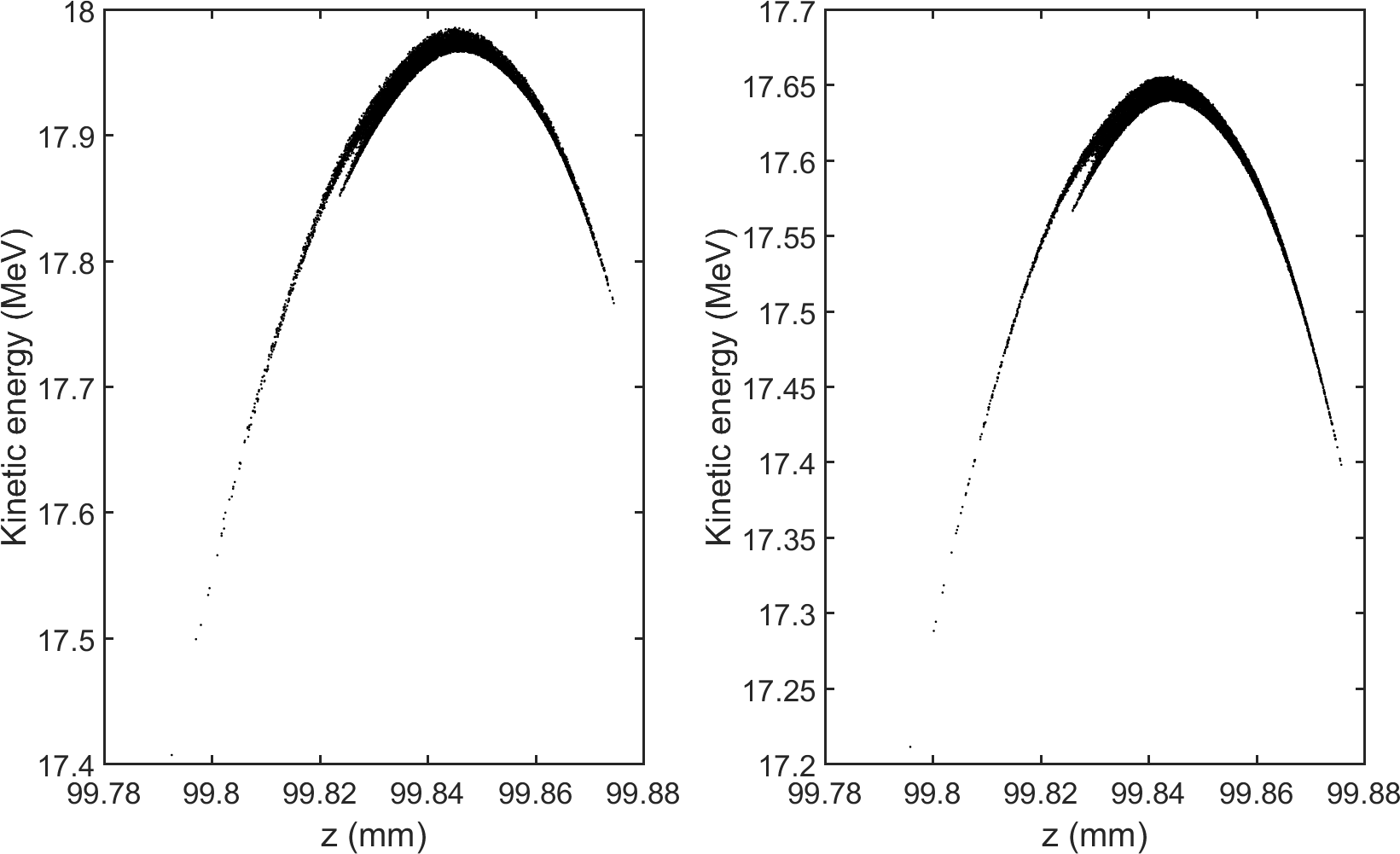}
  \caption{Longitudinal phase space of the bunch at the output for the tuned linac (left) and for the detuned linac with corrected frequency of the THz-pulse (right). }
      \label{Fig_5}
\end{figure}

As we are interested in the maximum output bunch kinetic energy, we compare the results of three simulations in order to define the influence of the fabrication errors. These simulations  correspond to the cases: (1) the optimized linac (with exact parameters), (2) the linac with 1.5 $\mu m$ outer dielectric radius fabrication error without the frequency correction (300 GHz) and (3) the linac with 1.5 $\mu m$ outer dielectric radius fabrication error with the frequency tune shift (296.696 GHz).  In order to find the maximum output bunch kinetic energy we use the 4.77 MeV point-like on-axis particles. Figure~\ref{Fig_3} shows results of these simulations.
Comparison of these three cases shows that by even small detuning of 1.5 $\mu m$ we will lose around 4 MeV bunch kinetic energy without frequency correction. But with frequency shift tune it is completely possible to compensate the detuning. More detailed analysis shows that the peaks on the curve of the output kinetic energy versus the injection time for the case of the detuned linac without the frequency correction become more narrow (as shown in zoomed parts of the Fig.~\ref{Fig_3}). As a consequence it will lead to an increase of the beam energy spread. After the simulations with the point-like beam we carried out acceleration simulations for the model beam for the cases of an ideally tuned linac and a detuned linac with corrected frequency. As for previous simulations the beam parameters at the linac entrance are set according to Table~\ref{Tabl_1}. In order to study the effect of the frequency correction for the detuned linac on the beam parameters we have compared the transverse and longitudinal phase spaces at the linac output. The results for the transverse phase space comparison including the Twiss parameters are presented in Fig.~\ref{Fig_4} (for both horizontal and vertical planes).
As it can be seen the transverse phase spaces of the bunch at the linac exit are very similar. In turn, from practical point of view it means that we do not need to change any beamline modules after the linac exit. We did the same comparison for the longitudinal phase space at the linac exit and the results are presented in Fig.~\ref{Fig_5}.
Figure~\ref{Fig_5} demonstrates that even with THz-pulse frequency tune there is little difference in the longitudinal phase spaces of the output beam. The reason is that the energy gain takes place as long as the bunch is in the accelerating phase of the THz-pulse~\cite{Galaydych_IPAC_2017}. The difference in maximum output energy is the difference in group velocities at 300 GHz for the tuned linac and for the detuned linac at 296.696 GHz, which is less than the speed of the electron bunch.

\begin{figure}[!h]
  \centering
  \includegraphics[width=\linewidth]{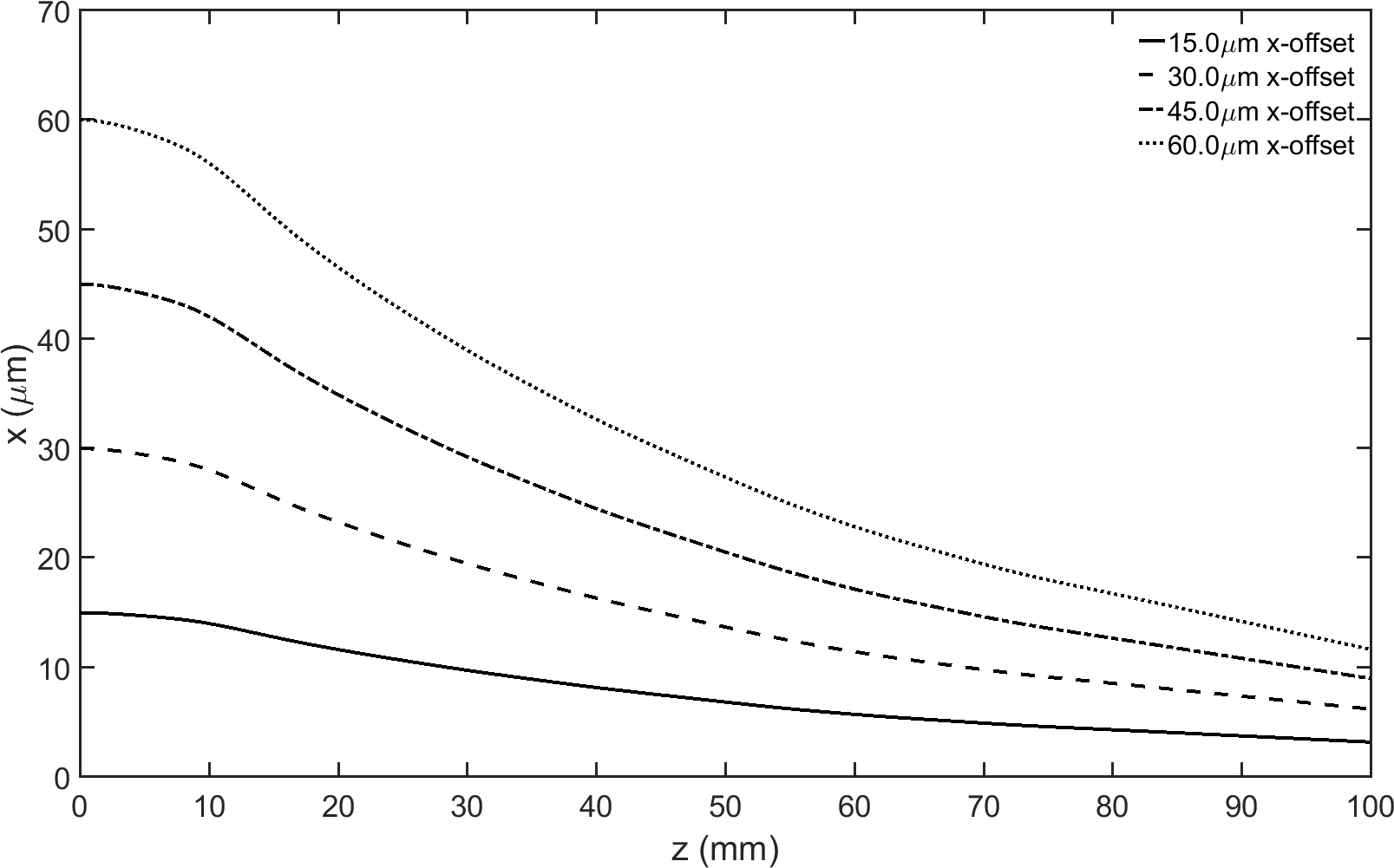}
  \caption{Evolution of the mean horizontal position of the beam during acceleration for the different values of the initial horizontal offsets: 15 $\mu m$ (solid line), 30 $\mu m$ (dashed line), 45 $\mu m$ (dash-dot line), 60 $\mu m$ (dotted line)}
      \label{Fig_6}
\end{figure}

Let us now consider beam injection transverse misalignments. As mentioned above, we will keep all the beam parameters as charge, average energy, energy spread and emittance the same. The first misalignment option is the horizontal (or vertical) offset of the injected beam. In this case we were mainly interested in radial stability of the acceleration. Figure~\ref{Fig_6} shows the dependence of the beam horizontal position on the longitudinal coordinate.

It can be seen that the correction of the radial beam position takes place during acceleration process. An explanation is that for the initial 4.77 MeV beam, in order to achieve the maximum energy at the linac output, we should inject the beam almost on crest. Since for the excited field $F_z\propto cos\phi$ and $F_r\propto -sin\phi$ (where $\phi$ is a phase of the beam relative to the field) ~\cite{Zhang}, there is an injection phase region, where simultaneous acceleration and focusing take place for the maximum energy gain regime. As a result due to rf focusing~\cite{Rosenzweig} the radial field will push the beam back to the axis. This means that for the chosen beam the initial offset is not a critical point for linac operation. We examined a range of beam offsets. Some of the output beam parameters are presented in Table~\ref{Tabl_2}. For offsets more than 60.0 $\mu m$ the beam starts to lose charge at the entrance aperture.
\begin{table}[!h]
  \centering
  \caption{Output parameters of the bunch for the different initial offsets.}
  \label{tab:table2}
  \begin{tabular}{cccc}
    \toprule
    Offset & $E_{kin}$ (MeV) & $\sigma_{x}$/$\sigma_{y}$ ($\mu m$) & $\epsilon_{x}$/$\epsilon_{y}$ ($\pi$ mm mrad)\\
    \midrule
    15 $\mu m$ & 17.4299 & 64.40 / 64.37 & 0.424 / 0.423\\
    30 $\mu m$ & 17.4301 & 64.31 / 64.33 & 0.422 / 0.423\\
    45 $\mu m$ & 17.4304 & 64.25 / 64.28 & 0.419 / 0.421\\
    60 $\mu m$ & 17.4308 & 64.26 / 64.24 & 0.418 / 0.420\\
    \bottomrule
  \end{tabular}
  \label{Tabl_2}
\end{table}

It should be noted that the transverse beam dynamics depend on the injection phase. As a result, for lower initial energy (for the maximal energy gain regime) the beam will deflect to the dielectric.

\begin{figure}[!h]
  \centering
  \includegraphics[width=\linewidth]{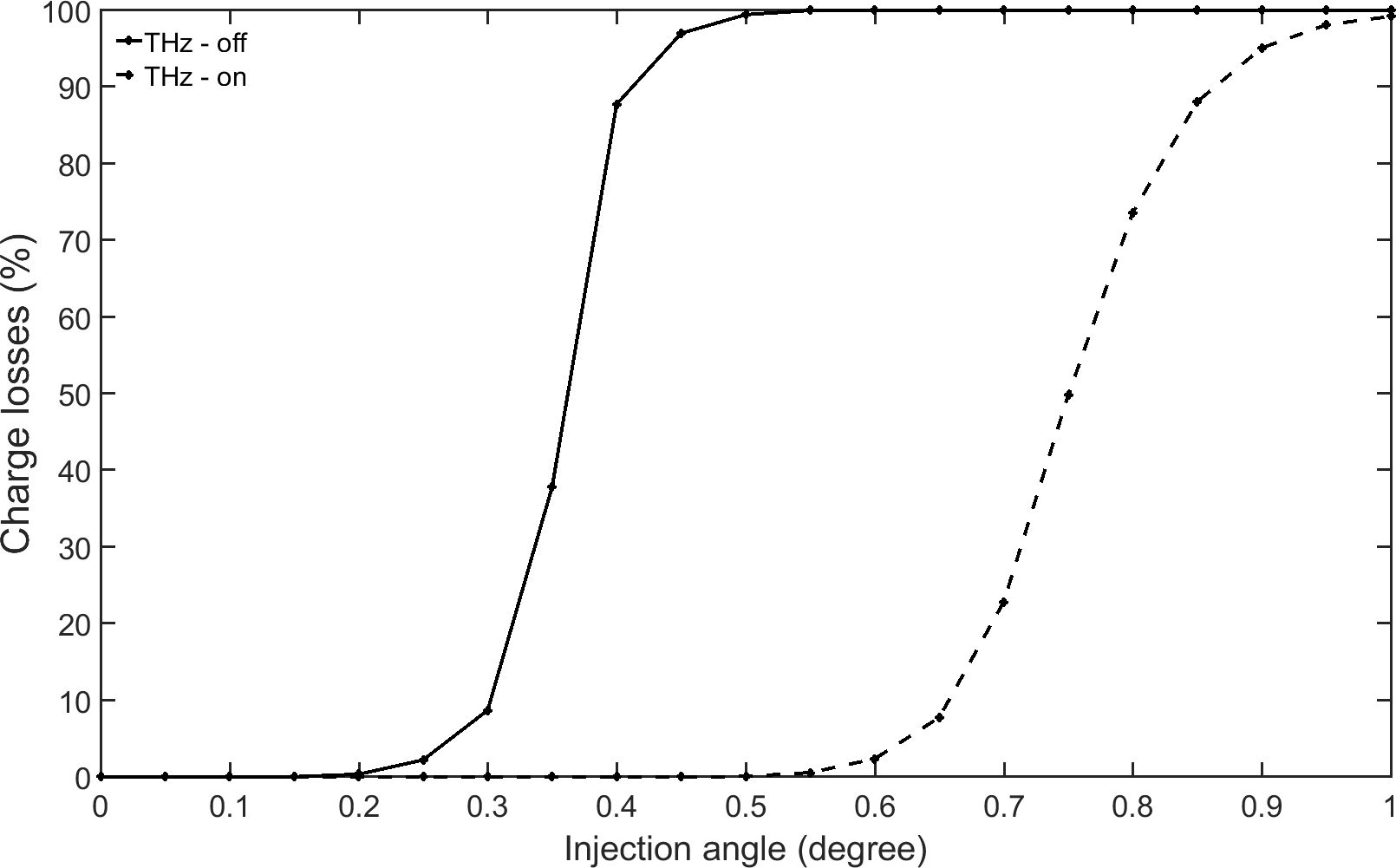}
  \caption{Charge losses as a function of the injection angle with respect to the longitudinal direction for two cases: (1, solid line) without the THz-pulse: linac geometrical acceptance, (2, dashed line) with THz-pulse.}
      \label{Fig_7}
\end{figure}

The second studied case of the beam misalignment is beam injection with non-zero azimuthal angle. For this type of injection we focused on the problem of beam losses on the dielectric surface, as the most critical one. Figure~\ref{Fig_7} illustrates the dependence of the charge losses on the angle of injection.
The obtained dependence shows that for a beam injection angle less than 0.5 degrees there are no charge losses at all and around 10\% (reasonable losses) of the charge is lost for an angle of 0.65 degrees. Additionally we have compared two cases: (1) with the THz-pulse and (2) without the THz-pulse. As it can be seen the injection angle at which charge losses reach around 10\% is about a factor of two larger for the case with THz-pulse. The reason is that due to the rf focusing, the THz-pulse corrects the beam to the linac axis. The accuracy of the necessary angle adjustment is technically realizable using stepper motor systems. It should be noted, that for the considered initial position and angle beam offsets (which lead to charge losses less than 10\%) output beam parameters stay almost the same.

\section{Conclusion}
This paper summarizes error and tolerance studies for the AXSIS linac. Requirements for the THz-pulse frequency tuning in order to compensate dielectric thickness fabrication errors were obtained. Simulation results demonstrate that the tuned linac and detuned linac with corrected THz-pulse frequency are almost the same. The beam offset studies show that for the initial 4.77 MeV electron beam and for the maximal energy gain regime the offset will be corrected by the field of the THz-pulse. Also it was demonstrated that injection angle misalignment, which lead to beam losses, can be adjusted by the linac alignment.

\section{Acknowledgment}

The authors would like to express special thanks to the AXSIS collaboration, and D. Marx for carefully proof reading the paper. This study is supported by the European Research Council under the European Union's Seventh Framework Programme (FP/2007-2013) / ERC Grant Agreement n. 609920.


\end{document}